\documentclass[aps,prd,preprint,superscriptaddress,nofootinbib,floatfix]{revtex4-2}
\usepackage[T1]{fontenc}
\usepackage[utf8]{inputenc}
\usepackage{amsmath,amssymb}
\usepackage{xcolor}
\usepackage{graphicx}
\usepackage{float}
\usepackage{placeins}
\usepackage{tikz}
\usepackage{pgfplots}
\pgfplotsset{compat=1.18}
\usepackage{calc}
\usepackage{longtable,booktabs,array}
\usepackage[hidelinks]{hyperref}
\makeatletter
\frontmatter@init
\makeatother

\begin{document}
\title{Big Mysteries Survey: Physicists' Views on Cosmology, Black Holes, Quantum Mechanics, and Quantum Gravity}
\author{Niayesh Afshordi}
\affiliation{Waterloo Centre for Astrophysics, University of Waterloo, 200 University Ave W, Waterloo, Ontario N2L 3G1, Canada}
\affiliation{Department of Physics and Astronomy, University of Waterloo, 200 University Ave W, Waterloo, Ontario N2L 3G1, Canada}
\affiliation{Perimeter Institute for Theoretical Physics, 31 Caroline St N, Waterloo, Ontario N2L 2Y5, Canada}
\author{Phil Halper}
\affiliation{Independent Scholar, London, UK}
\author{Matteo Rini}
\affiliation{Physics Magazine, American Physical Society}
\author{Michael Schirber}
\affiliation{Physics Magazine, American Physical Society}

\date{\today}

\begin{abstract}

We present results from the Big Mysteries Survey, a large-scale survey
conducted through the American Physical Society's \emph{Physics Magazine}
on foundational and controversial topics in contemporary physics. The
survey provides a snapshot of physicists' views on issues in cosmology,
black-hole physics, quantum mechanics, quantum gravity, and anthropic
coincidences. A central finding is that several positions
often described publicly as field-wide ``consensus'' views are, in
practice, supported by much narrower majorities or by pluralities rather
than majorities.
\end{abstract}

\maketitle

\section{Introduction}

In the summer of 2024, a survey was conducted at the Black Hole Inside
Out Conference in Copenhagen to assess physicists' views on a range of
ongoing controversies \cite{chen_halper_afshordi_2025}. Eighty-five scientists responded. One year
later, the authors collaborated with the American Physical Society's
\emph{Physics Magazine} on a substantially larger follow-up survey,
which polled 1,675 participants from the magazine's readership and the
members of the American Physical Society. The Physics Magazine survey
therefore provides a broader view of attitudes within the physics
community and allows comparisons with the more focused
conference-based Copenhagen sample.

Taken together, the two surveys make it possible to compare views
expressed in a specialist conference setting with those expressed by a
much larger and more heterogeneous respondent pool. On some topics, the
results are remarkably similar; on others, the differences are
substantial. This paper presents the Big Mysteries Survey results,
offers commentary on their interpretation, and highlights points of
agreement and divergence relative to the Copenhagen survey.

\begin{figure}[H]
\centering
\includegraphics[width=0.92\linewidth]{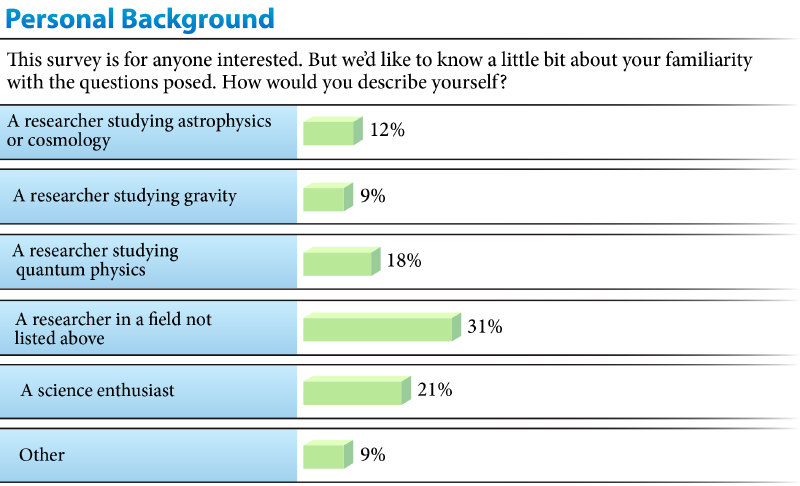}
\caption{Big Mysteries Survey (1,675 respondents): respondent background and self-reported profile information.}
\label{fig:aps-personal-background}
\end{figure}

To provide context for the interpretation of the response distributions,
Figure~\ref{fig:aps-personal-background} summarizes the respondent
background information supplied in the Physics Magazine survey. This
profile helps frame the results as reflecting a broad community sample
with mixed career stages and research backgrounds rather than a single
subdiscipline.

\section{Methods}

The Big Mysteries Survey was posted by \emph{Physics Magazine} from
28 July 2025 to 9 September 2025
\cite{aps_big_mysteries_survey_2025}. The questionnaire covered topics
in the foundations of quantum mechanics, cosmology, black-hole physics,
quantum gravity, and anthropic coincidences. In total, 1,675
participants responded and were asked to indicate their status within
the field.

Because the Physics Magazine survey was an open online survey rather
than a random sample of all physicists, the results should be interpreted as a
large-scale snapshot of respondents' views rather than as a precise
population estimate. Even with that caveat, the dataset is informative
because of its size and topical breadth. In what follows, we present the
response distributions and compare them, where relevant, with the
earlier Copenhagen conference survey \cite{chen_halper_afshordi_2025}.

For qualitative context, we also analyzed the free-text ``Other''
responses in the survey spreadsheet. These responses are included in
the anonymized dataset provided as supplemental material. They map in
order to Questions 1--10. For each question, we reviewed
all non-empty entries, grouped recurring themes by repeated concepts or
phrasing, and included short verbatim quotations as illustrative
examples in the corresponding Results subsection.

Finally, we study correlations among participant responses using a
log-odds statistic. The methodology is detailed in
Appendix~\ref{sec:correlation_methods}. In the cross-question analysis,
we exclude the trivial pairings (``Other'', ``Other'') and
(``No opinion'', ``No opinion''), which otherwise dominate the
highest-frequency combinations.

\section{Results and Discussion}

\subsection{Q1: What Does the Big Bang Imply?}\label{sec:q1}

\begin{figure}[H]
\centering
\includegraphics[width=0.92\linewidth]{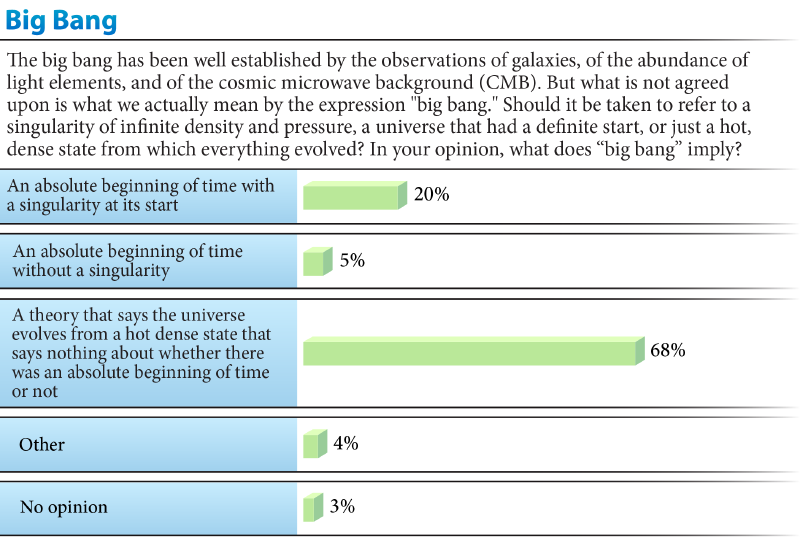}
\caption{Big Mysteries Survey (1,675 respondents): What does the expression "the big bang" imply?}
\label{fig:aps-bigbang-meaning}
\end{figure}

Multiple lines of evidence---including light-element abundances, the
large-scale distribution of galaxies, and the existence and thermal
properties of the cosmic microwave background (CMB)---support a Big Bang
expansion history. What remains contested is what physicists mean by the
phrase ``the Big Bang.'' Some leading cosmologists, most famously Stephen
Hawking~\cite{hawking2012beginning}, have argued that a beginning of
time is now effectively established. Others have challenged this
interpretation~\cite{afshordi_halper_2025}, emphasizing that the Big
Bang describes evolution from a hot, dense state and does not by itself
settle the question of whether time had an absolute beginning.

The Big Mysteries Survey responses strongly favor the latter interpretation
(Figure~\ref{fig:aps-bigbang-meaning}). Respondents overwhelmingly
prefer to treat the Big Bang as a theory of cosmic evolution from a hot,
dense state rather than as a statement that time began. In that sense,
it may be more accurate to say that physicists regard the universe as
\emph{at least} 13.8 billion years old, rather than to infer that they
thereby endorse a definite beginning of time.

Notably, this is the only question in the Big Mysteries Survey for which a single
option receives a large majority (more than two-thirds). In the Physics Magazine
sample, 68.4\% endorse the ``hot, dense state'' interpretation, very
close to the 68\% reported in the Copenhagen survey. A significant
difference between the two surveys, however, is the larger share of Physics Magazine
respondents who interpret the Big Bang as implying a singularity and a
beginning of time: 20\% in the Physics Magazine survey versus 11\% in
the Copenhagen survey. This does not necessarily imply belief in a
physically realized singularity; it
may instead reflect a difference in how respondents interpret the phrase
``Big Bang.''

The free-text ``Other'' responses for this question were dominated by
framing disputes over language rather than by a single alternative
theory. Representative comments include: ``This is a terminology
question. Different people use the term differently.'' Several other
comments similarly stressed that the meaning of ``Big Bang'' depends on
context.

\subsection{Q2: Early-Universe Puzzles and Inflation}\label{sec:q2}

\begin{figure}[H]
\centering
\includegraphics[width=0.92\linewidth]{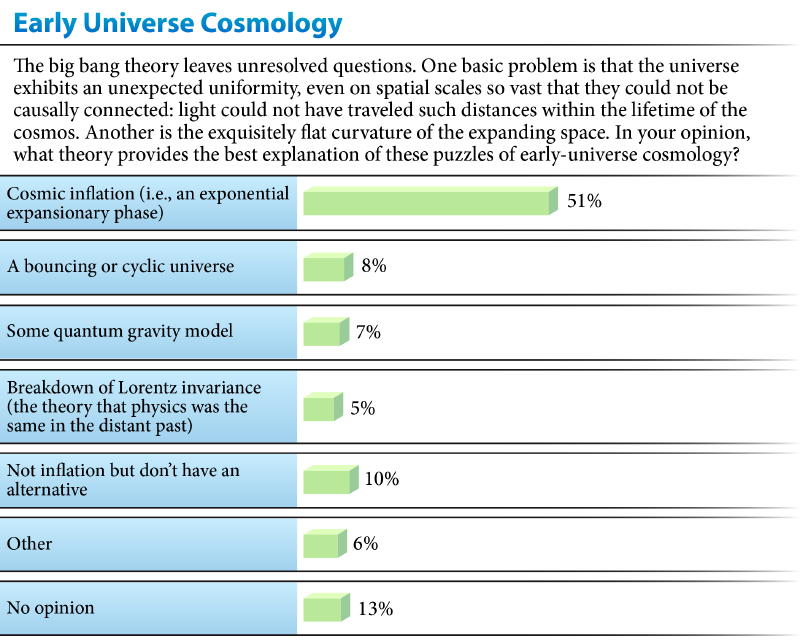}
\caption{Big Mysteries Survey (1,675 respondents): Which theory best explains early-universe puzzles such as horizon and flatness?}
\label{fig:aps-early-universe-puzzles}
\end{figure}

A second question addresses early-universe cosmology more directly by
asking what respondents regard as the most likely solution to the
horizon, structure, flatness, and monopole problems. Inflation is often
presented as the standard answer to these puzzles
\cite{guth1981inflationary}, but it has also faced sustained criticism
and competition from alternative proposals
\cite{ijjas_steinhardt_loeb_2013}.

The Big Mysteries Survey responses show no overwhelming support for any single
alternative, and inflation itself only narrowly exceeds the threshold
for majority support (Figure~\ref{fig:aps-early-universe-puzzles}). This
raises the question of whether inflation should be described as a strong
community consensus or, more cautiously, as the leading but contested
option. Compared with the Copenhagen survey, the pattern is broadly
similar, though inflation did not reach majority support there (44\%),
and ``some quantum gravity model'' was more popular (16\% in Copenhagen
versus 7\% in the Physics Magazine survey).

In the ``Other'' responses, the most common theme was hybrid early-universe
scenarios that combine mechanisms, for example: ``A combination of a
bouncing/cyclic universe AND quantum effects during the bounce.'' A
smaller but recurrent theme questioned whether the standard formulation
identifies a real paradox.

\subsection{Q3: Dark Matter, Modified Gravity, and Gravitational Anomalies}\label{sec:q3}

\begin{figure}[H]
\centering
\includegraphics[width=0.92\linewidth]{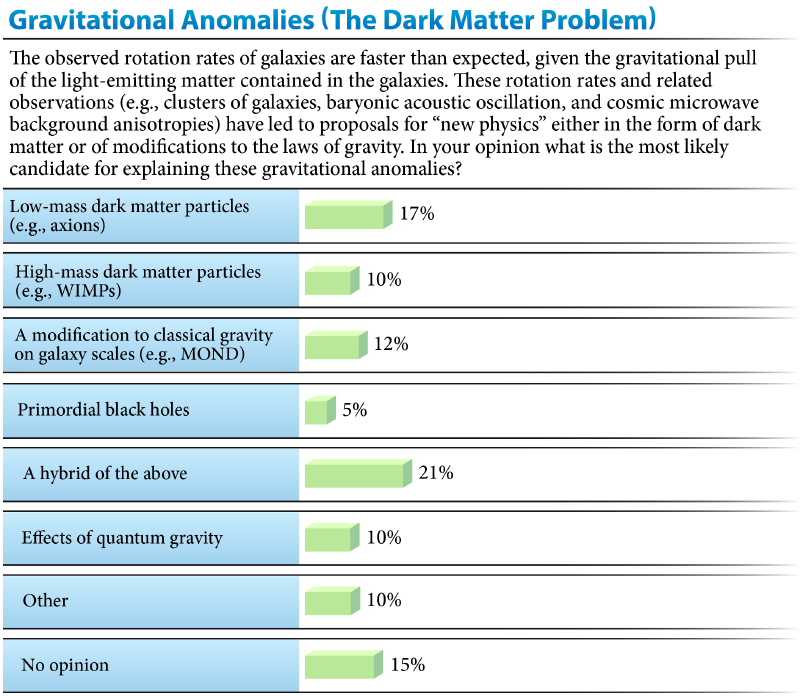}
\caption{Big Mysteries Survey (1,675 respondents): Most likely explanation for gravitational anomalies usually attributed to dark matter.}
\label{fig:aps-dark-matter-gravity-anomalies}
\end{figure}

Two survey questions address key components of the standard cosmological
model, $\Lambda$CDM, in which the universe is dominated by a
cosmological constant ($\Lambda$) and cold dark matter not accounted for
by the Standard Model of particle physics \cite{scott2020standard}. The
first asks about the most likely explanation of galactic rotation and
related gravitational anomalies.

No single dark-matter candidate dominates the Big Mysteries Survey
responses; instead, a hybrid answer is the most common choice
(Figure~\ref{fig:aps-dark-matter-gravity-anomalies}). One plausible
interpretation is that the non-detection of WIMPs in direct searches
\cite{arcadi2018waning}, together with the absence of supersymmetry
signals at the LHC \cite{atlas2022susy}, has reduced confidence in
WIMP-centered scenarios relative to other possibilities such as axions.

This overall pattern is consistent with the Copenhagen survey, where a
hybrid model was also the leading response. A notable difference,
however, is the stronger support in the Physics Magazine survey for
modifying gravity (or invoking quantum-gravity effects) to explain
phenomena often attributed to dark matter. In Copenhagen, modified
gravity and quantum gravity together received only 5\% support (1\%
MOND \cite{milgrom1983mond}, 4\% quantum gravity), whereas in the
Physics Magazine survey the corresponding numbers rise to 11.5\%
(MOND) and 10.1\% (quantum gravity). This increase appears to come
partly at the expense of primordial black holes, which received 17\%
support in Copenhagen but only 5.4\% in the Physics Magazine survey.
Even so, when all dark-matter-candidate options are combined (hybrid
models, axions, WIMPs, and primordial black holes), they exceed a
majority, totaling 53.4\%.

The ``Other'' free-text entries most often emphasized mixed or
nonexclusive explanations. Typical comments include ``I suspect that
Dark Matter and Dark Energy have a common explanation'' and
``Normal matter that we have either missed or miscounted!''
This pattern is consistent with the cross-question correlations below:
respondents selecting modified-gravity or quantum-gravity explanations
for dark-matter anomalies are also more likely to select related
responses for cosmic acceleration and the Hubble tension
(Table~\ref{tab:most-correlated}).

\subsection{Q4: Dark Energy and Cosmic Acceleration}\label{sec:q4}

\begin{figure}[H]
\centering
\includegraphics[width=0.92\linewidth]{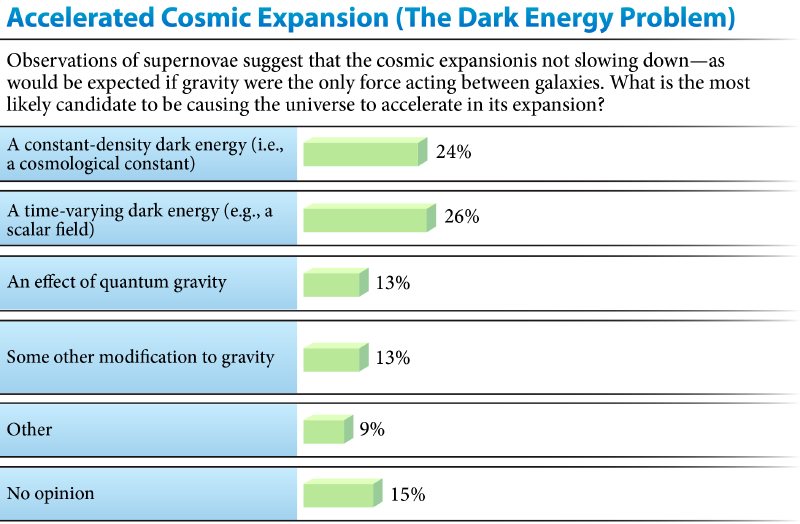}
\caption{Big Mysteries Survey (1,675 respondents): Most likely cause of the accelerated expansion of the universe.}
\label{fig:aps-cosmic-acceleration-cause}
\end{figure}

The second $\Lambda$CDM-related question concerns dark energy. Here the
Big Mysteries Survey responses are even less supportive of a canonical
textbook answer than in the dark-matter question. A time-varying field
is slightly more popular than a cosmological constant (25.9\% versus 24.0\%;
Figure~\ref{fig:aps-cosmic-acceleration-cause}). One possible
contributor to this shift is the influence of recent DESI-based analyses
\cite{lodha2025desi}, which have been widely discussed in connection
with evolving dark energy.

This represents a substantial change from the Copenhagen survey
(conducted before the latest DESI release), where a cosmological
constant was the most popular option (38\%) and a time-varying field
received only 16\% support. Importantly, however, neither survey finds a
majority for a cosmological constant. In that limited sense, many
physicists do not appear to endorse the simplest textbook formulation of
$\Lambda$CDM as strongly as public summaries sometimes imply.

The ``Other'' responses for this question were led by uncertainty and
model incompleteness, with many respondents explicitly stating that key
physics may still be missing. Examples include: ``We're missing some big
piece of physics that I hope we'll learn in my lifetime'' and ``With
current evidence from DESI I do not feel confident selecting a constant
or varying dark energy.''
In the correlation analysis, this question shows some of the strongest
cross-links in the survey: time-varying dark energy is tightly paired
with early dark energy in the Hubble-tension question, and quantum-gravity
answers here are tightly paired with quantum-gravity answers in both the
dark-matter and Hubble-tension questions
(Table~\ref{tab:most-correlated}).

\subsection{Q5: The Hubble Tension}\label{sec:q5}

\begin{figure}[H]
\centering
\includegraphics[width=0.92\linewidth]{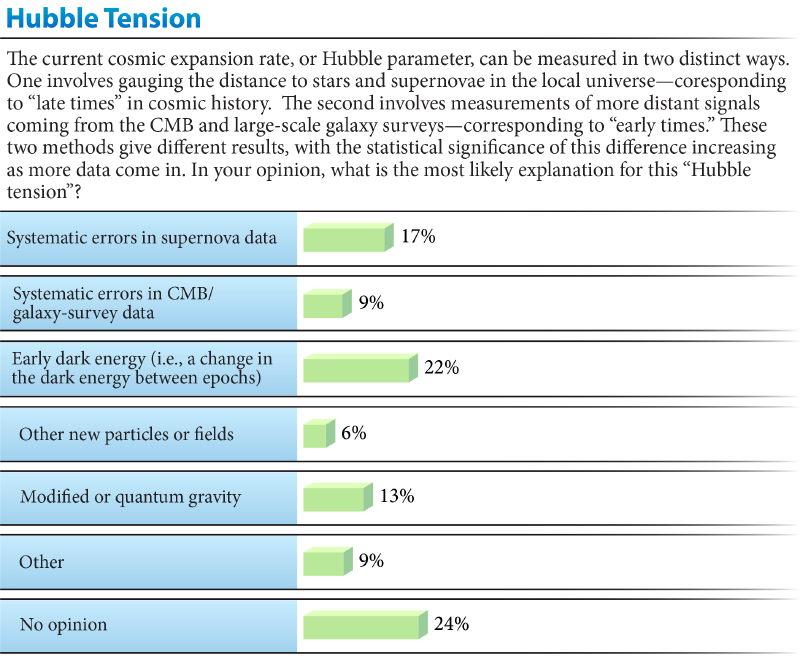}
\caption{Big Mysteries Survey (1,675 respondents): Most likely explanation for the Hubble tension.}
\label{fig:aps-hubble-tension}
\end{figure}

Another challenge to the standard cosmological picture is the Hubble
tension, in which two established methods of measuring the cosmic
expansion rate yield inconsistent values
\cite{kamionkowski_riess_2023,planck2020parameters,riess2022shoes}. Responses to this question indicate
less confidence than in many of the other topics surveyed. ``No
opinion'' is the most common response (24.4\%), followed by early dark
energy (22.1\%) (Figure~\ref{fig:aps-hubble-tension}).

This differs sharply from the Copenhagen survey, where systematic errors
in supernova data were the most popular explanation (35\%), compared
with only 16.7\% in the Physics Magazine survey. The contrast may
reflect differences in respondent composition, changes in the literature between the two
surveys, or both.

The dominant ``Other'' theme here was not a single new mechanism but
combined-systematics accounts, e.g., ``Both Systematic errors in
CMB/galaxy-survey data And Systematic errors in supernova data.'' A
second recurring theme was dissatisfaction with baseline modeling, such
as ``Problems with LCDM model used to model early universe.''
Consistent with this, correlated-response analysis indicates that many
respondents who endorse early dark energy for the Hubble tension also
endorse time-varying dark energy for cosmic acceleration, while those
choosing modified/quantum-gravity explanations for the Hubble tension
are strongly linked to analogous choices in the dark-matter and
dark-energy questions (Table~\ref{tab:most-correlated}).

\subsection{Q6: Anthropic Coincidences and Physical Constants}\label{sec:q6}

\begin{figure}[H]
\centering
\includegraphics[width=0.92\linewidth]{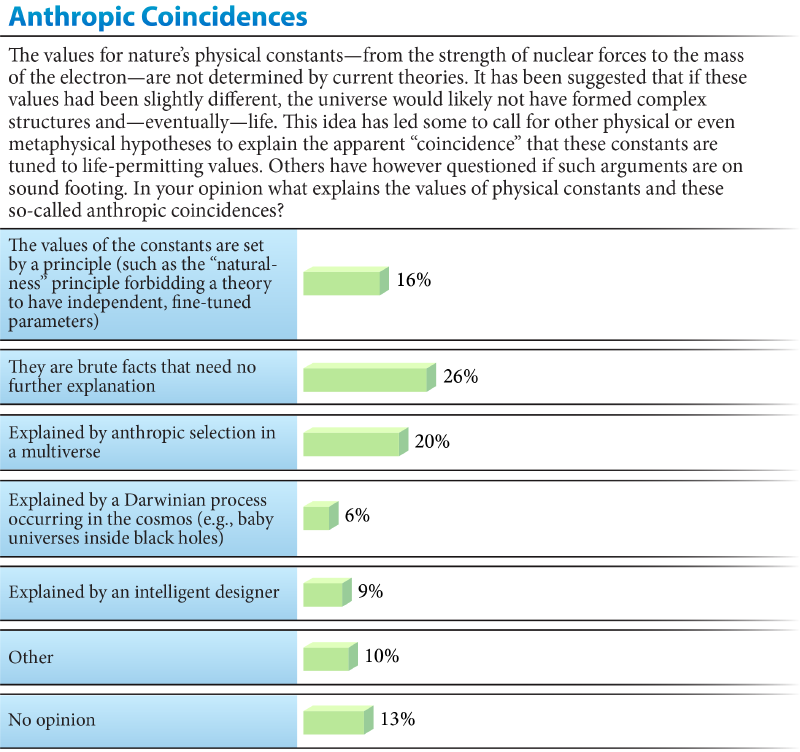}
\caption{Big Mysteries Survey (1,675 respondents): What explains anthropic coincidences and the observed values of physical constants?}
\label{fig:aps-anthropic-coincidences}
\end{figure}

It is often argued that small changes in the fundamental constants of
nature would prevent the emergence of complex structure or life
\cite{carter1974anthropic}. If so, what explains the observed
life-permitting values? Popular discussions frequently frame the issue
as forcing a choice between intelligent design and multiverse-based
anthropic selection \cite{barnes_lewis_2016}.

The Big Mysteries Survey results do not support that framing
(Figure~\ref{fig:aps-anthropic-coincidences}). The most common response
is that the constants are brute facts requiring no further explanation
(26\%). This was also the most common response in the Copenhagen survey
(33\%) \cite{chen_halper_afshordi_2025} and in the PhilPapers survey of philosophers (32\%) \cite{bourget_chalmers_2023}. Moreover,
even when the multiverse and intelligent-design responses are combined,
they still fall short of a majority (28.8\%). These results suggest that
many physicists do not accept the premise that fine-tuning arguments
compel commitment to either of those two explanatory frameworks.

The ``Other'' comments most commonly challenged narrow fine-tuning
framing and anthropocentric assumptions. Illustrative responses include:
``the anthropic principle means little'' and ``To think otherwise is
unforgivably anthropocentric.'' Another recurring theme asked for deeper
fundamental explanation rather than either design or multiverse
commitment.

\subsection{Q7: Interpretations of Quantum Mechanics}\label{sec:q7}

\begin{figure}[H]
\centering
\includegraphics[width=0.92\linewidth]{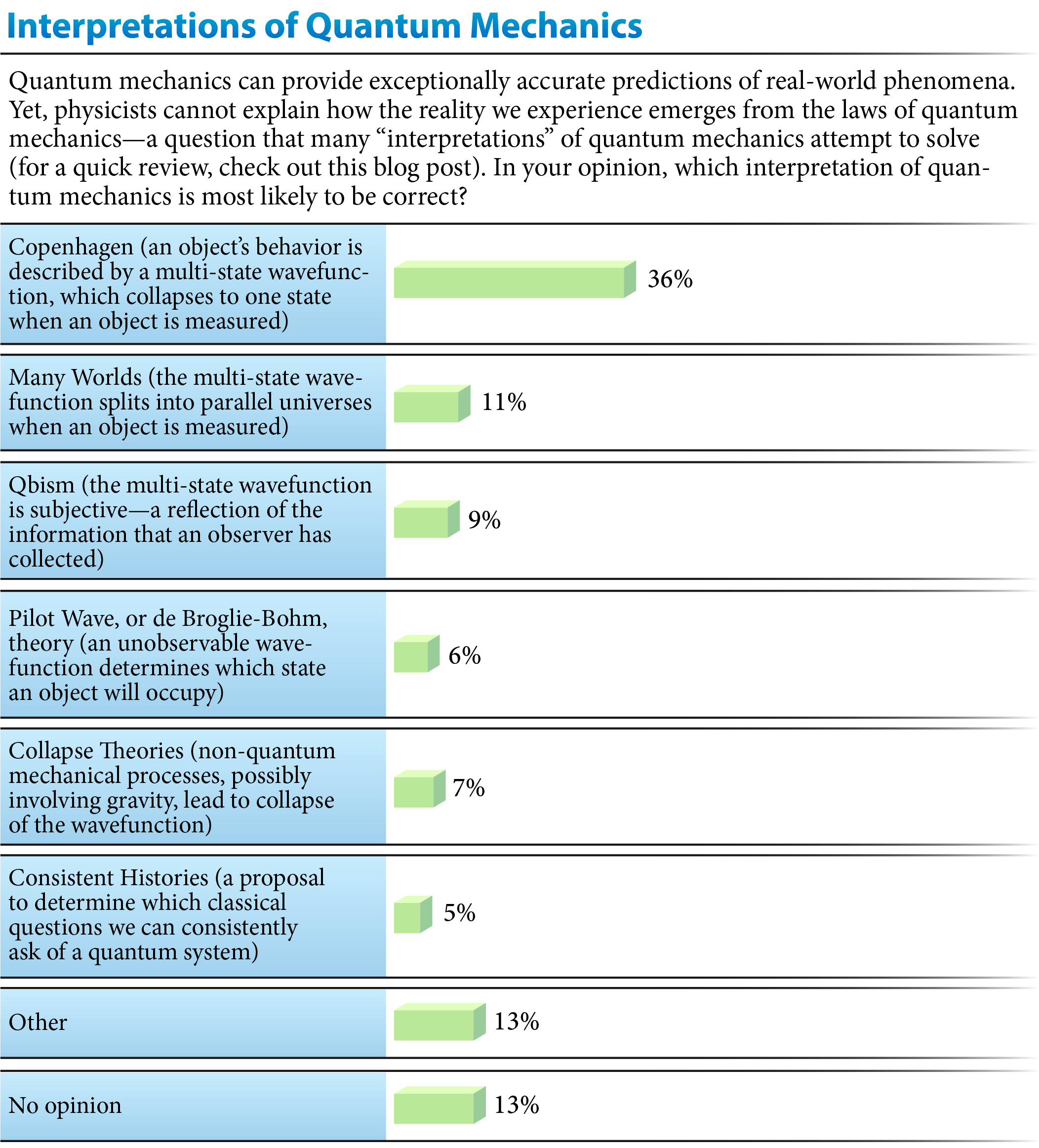}
\caption{Big Mysteries Survey (1,675 respondents): Which interpretation of quantum mechanics is most likely correct?}
\label{fig:aps-quantum-interpretations}
\end{figure}

Turning to quantum foundations, interpretations of quantum mechanics
have been the subject of several recent surveys. A widely discussed
\emph{Nature} survey~\cite{gibney2025quantum}, for example, found the
Copenhagen interpretation to be the most favored option (36\%). The
Physics Magazine survey yields a very similar top-line result, with
Copenhagen favored by 35.7\% of respondents
(Figure~\ref{fig:aps-quantum-interpretations}).

The ranking of alternatives differs, however. In the Physics Magazine
survey, Many-Worlds is the second most selected named interpretation
(11\%), whereas the \emph{Nature} survey reported epistemic
interpretations (such as Qbism) as the second most popular category
(17\%). It is also notable that, in the Physics Magazine sample, both
``no opinion'' and ``other'' outpoll Many-Worlds.
This underscores the extent of continuing fragmentation in views on the
interpretation of quantum mechanics. For canonical formulations or
modern introductions to several of the named options, see
\cite{everett1957,manyworlds_dewitt_graham_1973,bohm1952a,fuchs_mermin_schack_2014,ghirardi_rimini_weber_1986}.

In the ``Other'' free-text entries, the most common named alternatives
included relational quantum mechanics. A second recurring theme was
dissatisfaction with the provided taxonomy, captured by comments like
``None of the above.''
At the cross-question level, one notable high-log-odds pair links
``anthropic selection in a multiverse'' with ``Many Worlds,'' suggesting
that respondents attracted to explicitly multiverse-based reasoning in
cosmology are also more likely to select a multiverse-like quantum
interpretation (Table~\ref{tab:most-correlated}).

\subsection{Q8: Black Hole Event Horizons and the Fate of Infalling Matter}\label{sec:q8}

\begin{figure}[H]
\centering
\includegraphics[width=0.92\linewidth]{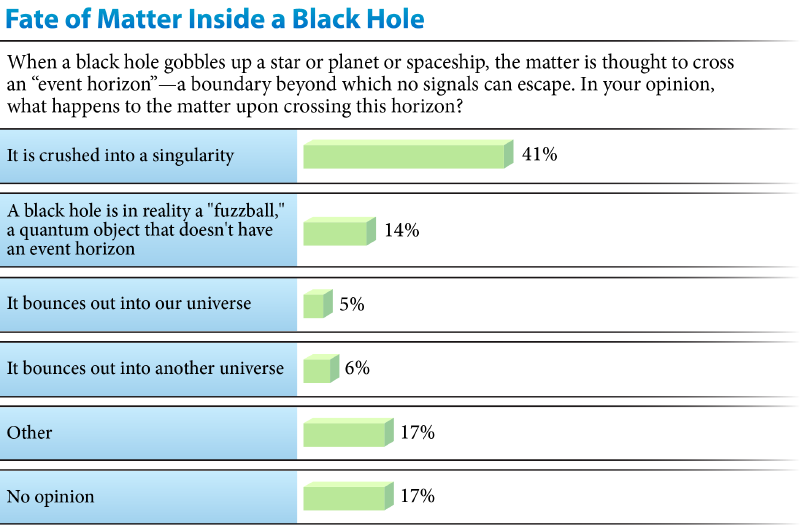}
\caption{Big Mysteries Survey (1,675 respondents): What happens to matter after crossing a black hole event horizon?}
\label{fig:aps-black-hole-horizon-matter}
\end{figure}

Another topic on which a textbook-style answer fails to command a
majority concerns the fate of matter crossing a black hole event
horizon. In standard general-relativistic language, infalling matter is
often said to end in a singularity \cite{senovilla2022penrose}. This is
the most popular Big Mysteries Survey response (40.5\%), but it still
falls well short of a majority
(Figure~\ref{fig:aps-black-hole-horizon-matter}).

This option is substantially more popular in the Physics Magazine survey
than in the Copenhagen survey (29\%). Conversely, alternatives such as
matter bouncing into our universe or another universe receive less
support in the Physics Magazine survey than in the Copenhagen sample.

Among ``Other'' responses, the most common theme was that horizon
crossing is locally uneventful: ``Nothing happens ... it simply continues
to free-fall forever.'' Closely related comments emphasized that crossing
does not immediately imply singular behavior: ``nothing special is
guaranteed to happen.''

\subsection{Q9: Black Hole Information}\label{sec:q9}

\begin{figure}[H]
\centering
\includegraphics[width=0.92\linewidth]{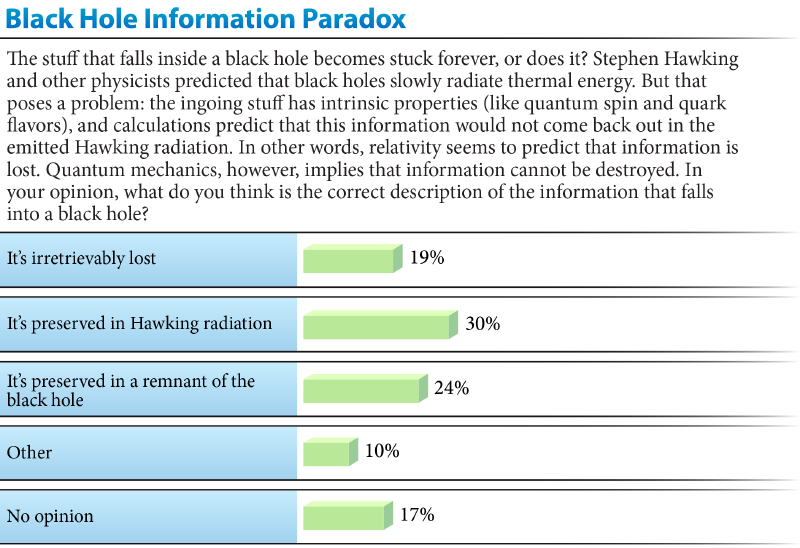}
\caption{Big Mysteries Survey (1,675 respondents): Which description of black hole information is most likely correct?}
\label{fig:aps-black-hole-information}
\end{figure}

Black holes also lie at the center of the information paradox. A common
way of stating the tension is that semiclassical gravity appears to
permit information loss, whereas quantum theory strongly motivates
information preservation \cite{mathur2009paradox}. It is sometimes
claimed that physicists now broadly agree that information is preserved,
with disagreement only about the mechanism
\cite{susskind2008blackholewar}. The Big Mysteries Survey results suggest that this
consensus narrative is overstated.

Information preservation only narrowly exceeds a majority in the
Physics Magazine survey. Combining preservation via Hawking radiation (30.5\%) and via
black hole remnants (23.7\%) yields 54.2\%
(Figure~\ref{fig:aps-black-hole-information}). Information loss remains
a minority view, but at 18.8\% it is far from negligible. These results
closely track the Copenhagen survey, which also found 53\% support for
information preservation (27\% Hawking radiation, 26\% remnants) and
18.8\% support for information loss.

In ``Other'' free-text comments, recurrent alternatives invoked
near-horizon structure and holography, for example: ``A black hole has
hair'' and ``Holographic event horizon.''
The correlation analysis adds one clear cross-domain association:
respondents who choose information preservation in Hawking radiation are
significantly more likely to select String Theory/M-Theory as their
preferred quantum-gravity framework
(Table~\ref{tab:most-correlated}).

\subsection{Q10: Candidates for Quantum Gravity}\label{sec:q10}

\begin{figure}[H]
\centering
\includegraphics[width=0.92\linewidth]{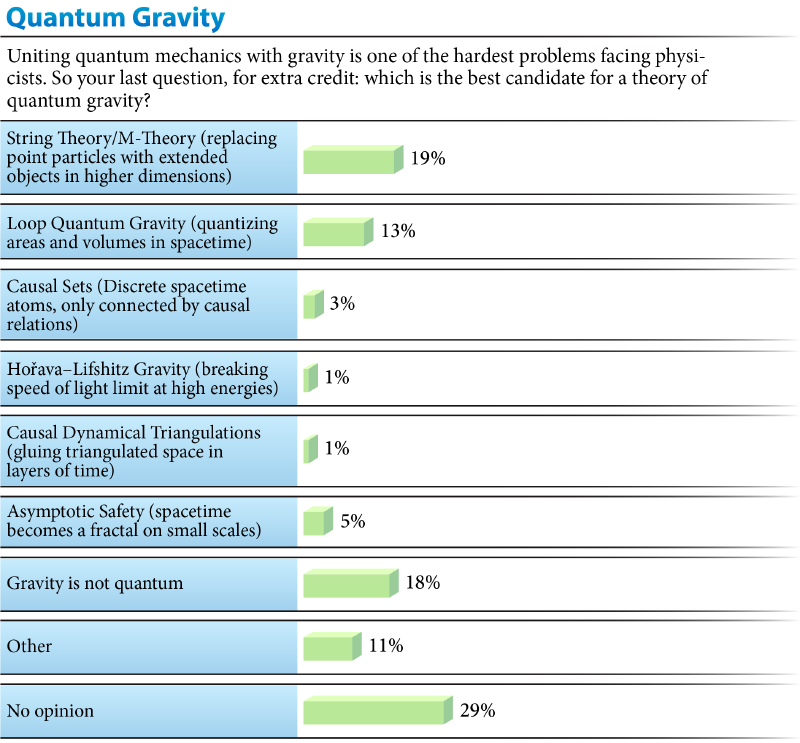}
\caption{Big Mysteries Survey (1,675 respondents): Which candidate is most likely to provide a theory of quantum gravity?}
\label{fig:aps-quantum-gravity-candidates}
\end{figure}

Perhaps the most prominent unresolved problem in fundamental physics is
the reconciliation of quantum mechanics with general relativity. The Big
Mysteries Survey responses show no dominant approach, and uncertainty is visible in the
fact that ``no opinion'' is the most selected response
(Figure~\ref{fig:aps-quantum-gravity-candidates}). String theory remains
the most popular named proposal, but with only 18.9\% support.

Notably, the view that gravity is not quantum ranks third (17.7\%),
ahead of loop quantum gravity (12.7\%). If one combines the listed
non-string quantum-gravity approaches, they account for 23.1\% of
responses, suggesting that even if string theory remains the best-known
candidate, many physicists do not treat it as uniquely compelling. The
overall pattern is similar to the Copenhagen survey, where string theory
received 21\% support and non-string alternatives combined received
19\%.

The ``Other'' responses most commonly expressed dissatisfaction with the
current menu of options, often simply ``none of the above.'' A related
theme argued that the correct framework is still missing, e.g., ``the
mathematical formula ... has not yet been developed.''

\subsection{Correlated Response Patterns Across Questions}

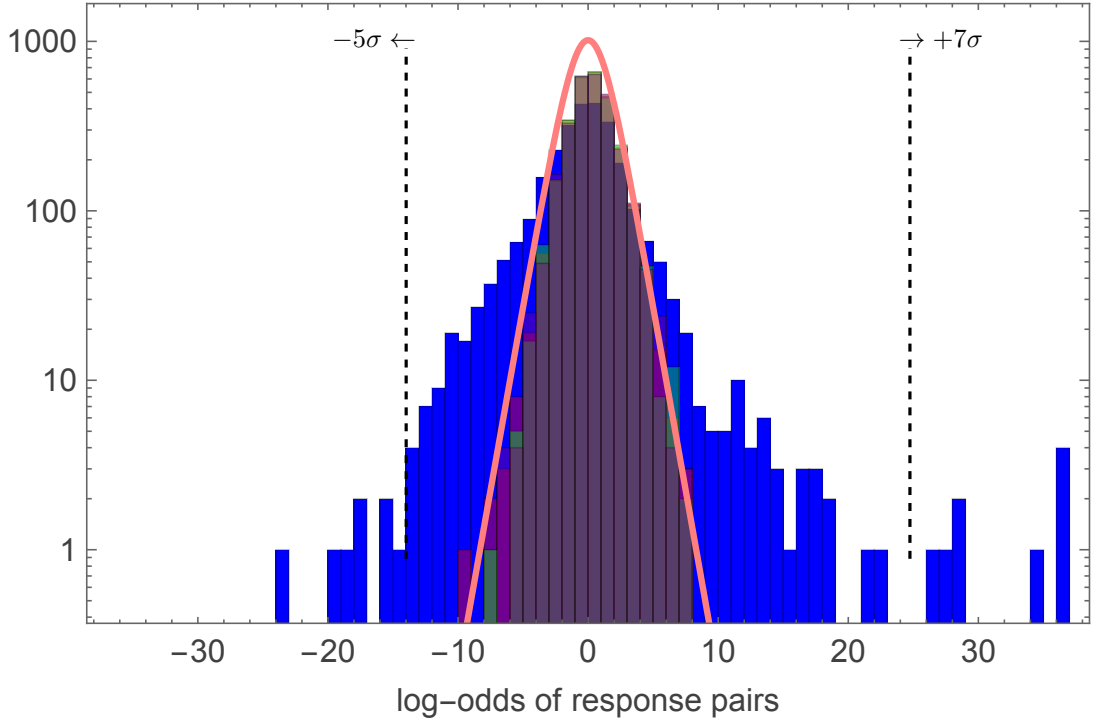
\begin{figure}[t]
\centering
\begin{tikzpicture}
\node[anchor=south west, inner sep=0] (img) at (0,0)
    {\includegraphics[width=0.92\linewidth]{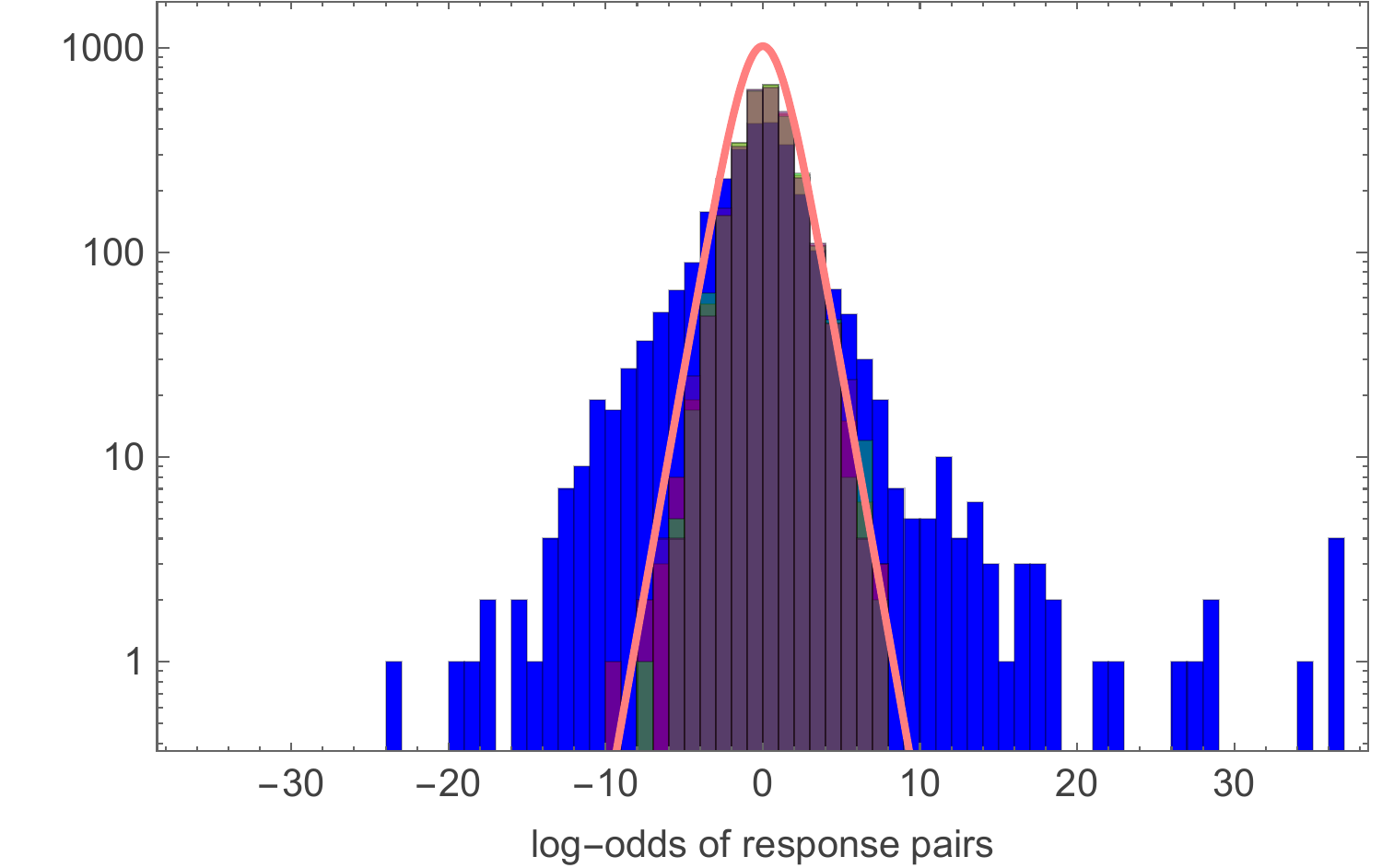}};
\begin{scope}[x={(img.south east)},y={(img.north west)}]
    \draw[very thick, dashed, black!70!black] (0.393,0.225) -- (0.393,0.935);
    \draw[very thick, dashed, black!80!black]  (0.833,0.225) -- (0.833,0.935);
    \node[anchor=south, text=black!70!black, font=\footnotesize] at (0.365,0.92) {$-5\sigma\leftarrow$};
    \node[anchor=south, text=black!80!black,  font=\footnotesize] at (0.86,0.92) {$\rightarrow+7\sigma$};
\end{scope}
\end{tikzpicture}
\caption{Distribution of log-odds values (blue histogram) for pairwise response correlations. Positive values indicate co-selection above chance expectation; negative values indicate co-selection below chance. The pink/red/purple histograms show three random realizations, assuming uncorrelated pairs, while the pink curve shows the theoretical prediction in Equation (\ref{eq:sech}). The overlaid dashed vertical lines mark the thresholds used in this paper: $\log\mathcal{O}<-15$ (approximately $-5\sigma$) for strongly anticorrelated pairs and $\log\mathcal{O}>25$ (approximately $+7\sigma$) for strongly correlated pairs.}
\label{fig:aps-logodds-map}
\end{figure}

Using the log-odds statistic defined in Appendix~\ref{sec:correlation_methods},
we identify highly correlated response pairs by $\log\mathcal{O}>25$
(roughly $+7\sigma$) and strongly anticorrelated pairs by
$\log\mathcal{O}<-15$ (roughly $-5\sigma$). The strongest positive
associations cluster around internally consistent theory bundles:
quantum-gravity explanations across early-universe, dark-matter, and
dark-energy questions; modified-gravity pairings across dark-matter and
dark-energy anomalies; and the dark-energy/Hubble-tension linkage
between time-varying dark energy and early dark energy. A separate
high-correlation cluster links black-hole information preservation in
Hawking radiation with string-theory preference for quantum gravity.
Figure~\ref{fig:aps-logodds-map} visualizes this split directly: the
right tail beyond the $+7\sigma$ line feeds Table~\ref{tab:most-correlated},
while the left tail beyond the $-5\sigma$ line feeds
Table~\ref{tab:least-correlated}.

\begin{table}[t]
\caption{Most positively correlated response pairs ($\log\mathcal{O}>25$).}
\label{tab:most-correlated}
\begin{tabular}{@{}p{0.68\linewidth}cc@{}}
\toprule
Pair of responses & $N_{\mathrm{both}}$ & Log-odds \\
\midrule
\hyperref[sec:q2]{Q2}: Some quantum gravity model $\leftrightarrow$ \hyperref[sec:q3]{Q3}: Effects of quantum gravity & 43 & 28.7 \\
\hyperref[sec:q2]{Q2}: Some quantum gravity model $\leftrightarrow$ \hyperref[sec:q5]{Q5}: Modified or quantum gravity & 49 & 26.7 \\
\hyperref[sec:q3]{Q3}: MOND/modification to gravity $\leftrightarrow$ \hyperref[sec:q4]{Q4}: Some other modification to gravity & 85 & 36.0 \\
\hyperref[sec:q3]{Q3}: Effects of quantum gravity $\leftrightarrow$ \hyperref[sec:q4]{Q4}: An effect of quantum gravity & 96 & 36.0 \\
\hyperref[sec:q3]{Q3}: Effects of quantum gravity $\leftrightarrow$ \hyperref[sec:q5]{Q5}: Modified or quantum gravity & 69 & 34.7 \\
\hyperref[sec:q4]{Q4}: Time-varying dark energy $\leftrightarrow$ \hyperref[sec:q5]{Q5}: Early dark energy & 219 & 36.0 \\
\hyperref[sec:q4]{Q4}: An effect of quantum gravity $\leftrightarrow$ \hyperref[sec:q5]{Q5}: Modified or quantum gravity & 88 & 36.0 \\
\hyperref[sec:q6]{Q6}: Anthropic selection in a multiverse $\leftrightarrow$ \hyperref[sec:q7]{Q7}: Many Worlds & 87 & 28.9 \\
\hyperref[sec:q9]{Q9}: Information preserved in Hawking radiation $\leftrightarrow$ \hyperref[sec:q10]{Q10}: String Theory/M-Theory & 172 & 28.0 \\
\bottomrule
\end{tabular}
\end{table}

\begin{table}[t]
\caption{Most anticorrelated response pairs ($\log\mathcal{O}<-15$).}
\label{tab:least-correlated}
\begin{tabular}{@{}p{0.68\linewidth}cc@{}}
\toprule
Pair of responses & $N_{\mathrm{both}}$ & Log-odds \\
\midrule
Personal background: Researcher studying gravity $\leftrightarrow$ \hyperref[sec:q10]{Q10}: No opinion & 12 & -18.0 \\
\hyperref[sec:q3]{Q3}: No opinion $\leftrightarrow$ \hyperref[sec:q5]{Q5}: Early dark energy & 21 & -16.0 \\
\hyperref[sec:q3]{Q3}: Other $\leftrightarrow$ \hyperref[sec:q8]{Q8}: Crushed into a singularity & 24 & -19.7 \\
\hyperref[sec:q4]{Q4}: Other $\leftrightarrow$ \hyperref[sec:q8]{Q8}: Crushed into a singularity & 26 & -15.5 \\
\hyperref[sec:q7]{Q7}: Other $\leftrightarrow$ \hyperref[sec:q8]{Q8}: Crushed into a singularity & 42 & -17.0 \\
\hyperref[sec:q8]{Q8}: Crushed into a singularity $\leftrightarrow$ \hyperref[sec:q9]{Q9}: Other & 23 & -23.1 \\
\hyperref[sec:q8]{Q8}: Crushed into a singularity $\leftrightarrow$ \hyperref[sec:q10]{Q10}: Other & 34 & -17.3 \\
\bottomrule
\end{tabular}
\end{table}

The negative-tail pairs mostly involve ``Other'' or ``No opinion''
responses, indicating that respondents who select specific mainstream
answers in one domain are statistically less likely than chance to
select broad residual categories in related domains.

\section{Conclusion}

In summary, the Big Mysteries Survey suggests that physicists are more
divided on many foundational questions than popular accounts often imply. Several
positions frequently presented as settled or near-consensus views do not
command majority support in this dataset. The clearest exception is the
interpretation of what the phrase ``Big Bang'' implies, for which more
than two-thirds of respondents prefer the ``hot, dense state'' reading
over an explicit beginning-of-time interpretation.

Across cosmology, the results indicate limited support for treating the
simplest textbook version of $\Lambda$CDM as an uncontroversial
consensus package, and they do not support the claim that fine-tuning
arguments force a choice between intelligent design and multiverse
anthropic selection. Inflation and black-hole information preservation
receive majority support, but only narrowly. In quantum foundations and
quantum gravity, the best-known proposals remain leading candidates, yet
none secures majority endorsement. More broadly, the survey highlights
the importance of distinguishing between ``most popular'' and
``consensus'' when describing contemporary physics to wider audiences.
The correlated-response analysis reinforces this picture by revealing
structured sub-communities of views rather than a single dominant
paradigm. The strongest positive pairs mostly connect conceptually
aligned answers across questions (for example, modified-gravity with
modified-gravity, quantum-gravity with quantum-gravity, and
time-varying dark energy with early dark energy), while the strongest
negative pairs are concentrated in ``Other''/``No opinion'' residual
categories. These correlations should be interpreted as patterns of
coherence in respondent worldviews, not as causal relationships between
the underlying physical hypotheses.

As seen in Figure~\ref{fig:aps-logodds-map}, the response patterns show
substantially richer correlation structure, and likely higher-order
organization, than we have analyzed in this study. We therefore provide
the anonymized raw data as supplemental material, so that other
researchers can further investigate this dataset, which offers a rare
window into how scientists think about major unresolved questions in
fundamental physics.

\begin{acknowledgments}
We are grateful to all survey participants for contributing their time
and perspectives. We also thank Carin Cain for the graphic
representations of the data. N.A. acknowledges support from the
University of Waterloo, the National Science and Engineering Research
Council of Canada (NSERC), and the Perimeter Institute for Theoretical
Physics. Research at Perimeter Institute is supported by the Government
of Canada through Industry Canada and by the Province of Ontario through
the Ministry of Economic Development and Innovation.

\end{acknowledgments}

\nocite{calmet2025entropy}
\FloatBarrier
\clearpage
\appendix
\onecolumngrid
\section{Statistical Measure of Correlation}\label{sec:correlation_methods}

In this appendix, we outline the statistical approach we use to study correlations amongst responses. 
Imagine the number of participants who choose the Response $R_1$ to the Question $Q_1$ is $n(R_1|Q_1)$. Similarly, the number of responses $R_2$ to Question $Q_2$ is $n(R_2|Q_2)$. Meanwhile, the number of people who respond $R_1$ to $Q_1$ {\it and} $R_2$ to $Q_2$ is $n(R_1,R_2|Q_1,Q_2)$. If there were no correlations amongst the responses, one would expect:
\begin{equation}
n_{\rm uncorr.} (R_1,R_2|Q_1,Q_2) = \frac{n(R_1|Q_1) \times n(R_2|Q_2)}{N}.
\end{equation}
We can now define the log-odds statistics as:
\begin{equation}
\log\mathcal{O}(R_1,R_2|Q_1,Q_2) \equiv \ln\left\{\frac{P\left[n(R_1,R_2|Q_1,Q_2) \leq n_{\rm uncorr.} (R_1,R_2|Q_1,Q_2)\right]}{P\left[n(R_1,R_2|Q_1,Q_2) >  n_{\rm uncorr.} (R_1,R_2|Q_1,Q_2)\right]}\right\},
\end{equation}
where $P$ is defined using cumulative Poisson probability distribution 
\begin{equation}
P[n \leq {\bar n}] = 1- P[n > {\bar n}] \equiv \frac{\Gamma(1+n,{\bar n})}{\Gamma(1+n)}.
\end{equation} 

If $n(R_1,R_2|Q_1,Q_2)$ comes from a random Poisson sampling of $n_{\rm uncorr.} (R_1,R_2|Q_1,Q_2)$, then it is easy to see that $\log\mathcal{O}$ should follow a probability distribution, centred around zero,  given by:
\begin{equation}
\frac{dP}{d\log\mathcal{O}}  = \frac{1}{4} {\rm sech}^2\left(\frac{\log\mathcal{O}}{2}\right).\label{eq:sech}
\end{equation}

Large positive (negative) values of log-odds that significantly deviate from this distribution suggest statistical correlation (anti-correlation) amongst responses by the participants. For examples, values of $|\log\mathcal{O}|>15$ ($27$) would correspond to 5$\sigma$  ($7\sigma$) anomalies from uncorrelated random distribution.

\section{Full Survey Question Tables}\label{app:full-survey-tables}
This appendix reproduces the full question wording and full answer choices shown in the original survey table images, along with the reported counts and percentages.
Uncertainties are reported as Poisson 1$\sigma$ errors, with $\sigma_n=\sqrt{n}$ for counts and propagated percentage errors $\sigma_f = 100\,\sqrt{n}/1675$.

\setlength{\LTleft}{0pt}
\setlength{\LTright}{0pt}
\subsection{Question 1: Big Bang Meaning}
\noindent\textbf{Question.} The big bang has been well established by the observations of galaxies, of the abundance of light elements, and of the cosmic microwave background (CMB). But what is not agreed upon is what we actually mean by the phrase big bang. Should it be taken to refer to a singularity of infinite density and pressure, a universe that had a definite start, or just a hot, dense state from which everything evolved? In your opinion, what does the expression ``the big bang'' imply?

\begin{longtable}{@{}p{0.66\textwidth}r r@{}}
\toprule
Answer & Count ($\pm$1$\sigma$) & Percent ($\pm$1$\sigma$) \\
\midrule
\endfirsthead
\toprule
Answer & Count ($\pm$1$\sigma$) & Percent ($\pm$1$\sigma$) \\
\midrule
\endhead
\bottomrule
\endfoot
An absolute beginning of time with a singularity at its start & 325 $\pm$ 18.0 & 19.5 $\pm$ 1.1 \\
An absolute beginning of time without a singularity & 75 $\pm$ 8.7 & 4.5 $\pm$ 0.5 \\
A theory that says the universe evolves from a hot dense state that says nothing about whether there was an absolute beginning of time or not & 1141 $\pm$ 33.8 & 68.4 $\pm$ 2.0 \\
Other & 69 $\pm$ 8.3 & 4.1 $\pm$ 0.5 \\
No opinion & 57 $\pm$ 7.5 & 3.4 $\pm$ 0.5 \\
\end{longtable}

\subsection{Question 2: Early-Universe Puzzles}
\noindent\textbf{Question.} The big bang theory leaves unresolved questions. One basic problem is that the universe exhibits an unexpected uniformity, even on spatial scales so vast that they could not be causally connected: light could not have traveled such distances within the lifetime of the cosmos. Another is the exquisitely flat curvature of the expanding space. In your opinion, what theory provides the best explanation of these puzzles of early-universe cosmology?

\begin{longtable}{@{}p{0.66\textwidth}r r@{}}
\toprule
Answer & Count ($\pm$1$\sigma$) & Percent ($\pm$1$\sigma$) \\
\midrule
\endfirsthead
\toprule
Answer & Count ($\pm$1$\sigma$) & Percent ($\pm$1$\sigma$) \\
\midrule
\endhead
\bottomrule
\endfoot
Cosmic inflation (i.e., an exponential expansionary phase) & 847 $\pm$ 29.1 & 50.8 $\pm$ 1.7 \\
A bouncing or cyclic universe & 129 $\pm$ 11.4 & 7.7 $\pm$ 0.7 \\
Some quantum gravity model & 116 $\pm$ 10.8 & 7.0 $\pm$ 0.6 \\
Breakdown of Lorentz invariance (the theory that physics was the same in the distant past) & 87 $\pm$ 9.3 & 5.2 $\pm$ 0.6 \\
Not inflation but don't have an alternative & 169 $\pm$ 13.0 & 10.1 $\pm$ 0.8 \\
Other & 94 $\pm$ 9.7 & 5.6 $\pm$ 0.6 \\
No opinion & 226 $\pm$ 15.0 & 13.5 $\pm$ 0.9 \\
\end{longtable}

\subsection{Question 3: Dark Matter or Modified Gravity}
\noindent\textbf{Question.} The observed rotation rates of galaxies are faster than expected, given the gravitational pull of the light-emitting matter contained in the galaxies. These rotation rates and related observations (e.g., clusters of galaxies, baryonic acoustic oscillation, and cosmic microwave background anisotropies) have led to proposals for ``new physics'' either in the form of dark matter or of modifications to the laws of gravity. In your opinion what is the most likely candidate for explaining these gravitational anomalies?

\begin{longtable}{@{}p{0.66\textwidth}r r@{}}
\toprule
Answer & Count ($\pm$1$\sigma$) & Percent ($\pm$1$\sigma$) \\
\midrule
\endfirsthead
\toprule
Answer & Count ($\pm$1$\sigma$) & Percent ($\pm$1$\sigma$) \\
\midrule
\endhead
\bottomrule
\endfoot
Low-mass dark matter particles (e.g., axions) & 290 $\pm$ 17.0 & 17.4 $\pm$ 1.0 \\
High-mass dark matter particles (e.g., WIMPs) & 167 $\pm$ 12.9 & 10.0 $\pm$ 0.8 \\
A modification to classical gravity on galaxy scales (e.g., MOND) & 192 $\pm$ 13.9 & 11.5 $\pm$ 0.8 \\
Primordial black holes & 90 $\pm$ 9.5 & 5.4 $\pm$ 0.6 \\
A hybrid of the above & 344 $\pm$ 18.5 & 20.6 $\pm$ 1.1 \\
Effects of quantum gravity & 168 $\pm$ 13.0 & 10.1 $\pm$ 0.8 \\
Other & 164 $\pm$ 12.8 & 9.8 $\pm$ 0.8 \\
No opinion & 252 $\pm$ 15.9 & 15.1 $\pm$ 0.9 \\
\end{longtable}

\subsection{Question 4: Cause of Cosmic Acceleration}
\noindent\textbf{Question.} Observations of supernovae suggest that the cosmic expansion is not slowing down---as would be expected if gravity were the only force acting between galaxies. What is the most likely candidate to be causing the universe to accelerate in its expansion?

\begin{longtable}{@{}p{0.66\textwidth}r r@{}}
\toprule
Answer & Count ($\pm$1$\sigma$) & Percent ($\pm$1$\sigma$) \\
\midrule
\endfirsthead
\toprule
Answer & Count ($\pm$1$\sigma$) & Percent ($\pm$1$\sigma$) \\
\midrule
\endhead
\bottomrule
\endfoot
A constant-density dark energy (i.e., a cosmological constant) & 400 $\pm$ 20.0 & 24.0 $\pm$ 1.2 \\
A time-varying dark energy (e.g., a scalar field) & 431 $\pm$ 20.8 & 25.9 $\pm$ 1.2 \\
An effect of quantum gravity & 214 $\pm$ 14.6 & 12.9 $\pm$ 0.9 \\
Some other modification to gravity & 216 $\pm$ 14.7 & 13.0 $\pm$ 0.9 \\
Other & 155 $\pm$ 12.4 & 9.3 $\pm$ 0.7 \\
No opinion & 249 $\pm$ 15.8 & 15.0 $\pm$ 0.9 \\
\end{longtable}

\subsection{Question 5: Hubble Tension}
\noindent\textbf{Question.} The current cosmic expansion rate, or Hubble parameter, can be measured in two distinct ways. One involves gauging the distance to stars and supernovae in the local universe---corresponding to ``late times'' in cosmic history. The second involves measurements of more distant signals coming from the CMB and large-scale galaxy surveys---corresponding to ``early times.'' These two methods give different results, with the statistical significance of this difference increasing as more data come in. In your opinion, what is the most likely explanation for this ``Hubble tension''?

\begin{longtable}{@{}p{0.66\textwidth}r r@{}}
\toprule
Answer & Count ($\pm$1$\sigma$) & Percent ($\pm$1$\sigma$) \\
\midrule
\endfirsthead
\toprule
Answer & Count ($\pm$1$\sigma$) & Percent ($\pm$1$\sigma$) \\
\midrule
\endhead
\bottomrule
\endfoot
Systematic errors in supernova data & 277 $\pm$ 16.6 & 16.7 $\pm$ 1.0 \\
Systematic errors in CMB/galaxy-survey data & 146 $\pm$ 12.1 & 8.8 $\pm$ 0.7 \\
Early dark energy (i.e., a change in the dark energy between epochs) & 368 $\pm$ 19.2 & 22.1 $\pm$ 1.1 \\
Other new particles or fields & 92 $\pm$ 9.6 & 5.5 $\pm$ 0.6 \\
Modified or quantum gravity & 223 $\pm$ 14.9 & 13.4 $\pm$ 0.9 \\
Other & 150 $\pm$ 12.2 & 9.0 $\pm$ 0.7 \\
No Opinion & 406 $\pm$ 20.1 & 24.4 $\pm$ 1.2 \\
\end{longtable}

\subsection{Question 6: Anthropic Coincidences}
\noindent\textbf{Question.} The values for nature's physical constants---from the strength of nuclear forces to the mass of the electron---are not determined by current theories, It has been suggested that if these values had been slightly different, the universe would likely not have formed complex structures and---eventually---life. This idea has led some to call for other physical or even metaphysical hypotheses to explain the apparent ``coincidence'' that these constants are tuned to life-permitting values. Others have however questioned if such arguments are on sound footing. In your opinion what explains the values of physical constants and these so-called anthropic coincidences?

\begin{longtable}{@{}p{0.66\textwidth}r r@{}}
\toprule
Answer & Count ($\pm$1$\sigma$) & Percent ($\pm$1$\sigma$) \\
\midrule
\endfirsthead
\toprule
Answer & Count ($\pm$1$\sigma$) & Percent ($\pm$1$\sigma$) \\
\midrule
\endhead
\bottomrule
\endfoot
The values of the constants are set by a principle (such as the ``naturalness'' principle forbidding a theory to have independent, fine-tuned parameters) & 261 $\pm$ 16.2 & 15.7 $\pm$ 1.0 \\
They are brute facts that need no further explanation & 433 $\pm$ 20.8 & 26.0 $\pm$ 1.2 \\
Explained by anthropic selection in a multiverse & 332 $\pm$ 18.2 & 19.9 $\pm$ 1.1 \\
Explained by a Darwinian process occurring in the cosmos (e.g., baby universes inside black holes) & 104 $\pm$ 10.2 & 6.2 $\pm$ 0.6 \\
Explained by an intelligent designer & 148 $\pm$ 12.2 & 8.9 $\pm$ 0.7 \\
Other & 177 $\pm$ 13.3 & 10.6 $\pm$ 0.8 \\
No opinion & 210 $\pm$ 14.5 & 12.6 $\pm$ 0.9 \\
\end{longtable}

\subsection{Question 7: Quantum Mechanics Interpretations}
\noindent\textbf{Question.} Quantum mechanics can provide exceptionally accurate predictions of real-world phenomena. Yet, physicists cannot explain how the reality we experience emerges from the laws of quantum mechanics---a question that many ``interpretations'' of quantum mechanics attempt to solve (for a quick review, check out this blog post). In your opinion, which interpretation of quantum mechanics is most likely to be correct?

\begin{longtable}{@{}p{0.66\textwidth}r r@{}}
\toprule
Answer & Count ($\pm$1$\sigma$) & Percent ($\pm$1$\sigma$) \\
\midrule
\endfirsthead
\toprule
Answer & Count ($\pm$1$\sigma$) & Percent ($\pm$1$\sigma$) \\
\midrule
\endhead
\bottomrule
\endfoot
Copenhagen (an object's behavior is described by a multi-state wavefunction, which collapses to one state when an object is measured) & 594 $\pm$ 24.4 & 35.7 $\pm$ 1.5 \\
Many Worlds (the multi-state wavefunction splits into parallel universes when an object is measured) & 183 $\pm$ 13.5 & 11.0 $\pm$ 0.8 \\
Qbism (the multi-state wavefunction is subjective---a reflection of the information that an observer has collected) & 153 $\pm$ 12.4 & 9.2 $\pm$ 0.7 \\
Pilot Wave, or de Broglie--Bohm, theory (an unobservable wavefunction determines which state an object will occupy) & 96 $\pm$ 9.8 & 5.8 $\pm$ 0.6 \\
Collapse Theories (non-quantum mechanical processes, possibly involving gravity, lead to collapse of the wavefunction) & 108 $\pm$ 10.4 & 6.5 $\pm$ 0.6 \\
Consistent Histories (a proposal to determine which classical questions we can consistently ask of a quantum system) & 87 $\pm$ 9.3 & 5.2 $\pm$ 0.6 \\
Other & 219 $\pm$ 14.8 & 13.2 $\pm$ 0.9 \\
No opinion & 223 $\pm$ 14.9 & 13.4 $\pm$ 0.9 \\
\end{longtable}

\subsection{Question 8: Matter Crossing an Event Horizon}
\noindent\textbf{Question.} When a black hole gobbles up a star or planet or spaceship, the matter is thought to cross an ``event horizon''---a boundary beyond which no signals can escape. In your opinion, what happens to the matter upon crossing this horizon?

\begin{longtable}{@{}p{0.66\textwidth}r r@{}}
\toprule
Answer & Count ($\pm$1$\sigma$) & Percent ($\pm$1$\sigma$) \\
\midrule
\endfirsthead
\toprule
Answer & Count ($\pm$1$\sigma$) & Percent ($\pm$1$\sigma$) \\
\midrule
\endhead
\bottomrule
\endfoot
It is crushed into a singularity & 672 $\pm$ 25.9 & 40.5 $\pm$ 1.5 \\
A black hole is in reality a "fuzzball," a quantum object that doesn't have an event horizon. & 237 $\pm$ 15.4 & 14.3 $\pm$ 0.9 \\
It bounces out into our universe & 81 $\pm$ 9.0 & 4.9 $\pm$ 0.5 \\
It bounces out into another universe & 106 $\pm$ 10.3 & 6.4 $\pm$ 0.6 \\
Other & 275 $\pm$ 16.6 & 16.6 $\pm$ 1.0 \\
No opinion & 287 $\pm$ 16.9 & 17.3 $\pm$ 1.0 \\
\end{longtable}

\subsection{Question 9: Black Hole Information}
\noindent\textbf{Question.} The stuff that falls inside a black hole becomes stuck forever, or does it? Stephen Hawking and other physicists predicted that black holes slowly radiate thermal energy. But that poses a problem: the ingoing stuff has intrinsic properties (like quantum spin and quark flavors), and calculations predict that this information would not come back out in the emitted Hawking radiation. In other words, relativity seems to predict information is lost. Quantum mechanics, however, implies that information cannot be destroyed. In your opinion, what do you think is the correct description of the information that falls into a black hole?

\begin{longtable}{@{}p{0.66\textwidth}r r@{}}
\toprule
Answer & Count ($\pm$1$\sigma$) & Percent ($\pm$1$\sigma$) \\
\midrule
\endfirsthead
\toprule
Answer & Count ($\pm$1$\sigma$) & Percent ($\pm$1$\sigma$) \\
\midrule
\endhead
\bottomrule
\endfoot
It's irretrievably lost & 312 $\pm$ 17.7 & 18.8 $\pm$ 1.1 \\
It's preserved in Hawking radiation & 508 $\pm$ 22.5 & 30.5 $\pm$ 1.3 \\
It's preserved in a remnant of the black hole & 394 $\pm$ 19.8 & 23.7 $\pm$ 1.2 \\
Other & 173 $\pm$ 13.2 & 10.4 $\pm$ 0.8 \\
No opinion & 276 $\pm$ 16.6 & 16.6 $\pm$ 1.0 \\
\end{longtable}

\subsection{Question 10: Quantum Gravity Candidates}
\noindent\textbf{Question.} Uniting quantum mechanics with gravity is one of the hardest problems facing physicists. So your last question, for extra credit: which is the best candidate for a theory of quantum gravity?

\begin{longtable}{@{}p{0.66\textwidth}r r@{}}
\toprule
Answer & Count ($\pm$1$\sigma$) & Percent ($\pm$1$\sigma$) \\
\midrule
\endfirsthead
\toprule
Answer & Count ($\pm$1$\sigma$) & Percent ($\pm$1$\sigma$) \\
\midrule
\endhead
\bottomrule
\endfoot
String Theory/M-Theory (replacing point particles with extended objects in higher dimensions) & 315 $\pm$ 17.7 & 18.9 $\pm$ 1.1 \\
Loop Quantum Gravity (quantizing areas and volumes in spacetime) & 212 $\pm$ 14.6 & 12.7 $\pm$ 0.9 \\
Causal Sets (Discrete spacetime atoms, only connected by causal relations) & 47 $\pm$ 6.9 & 2.8 $\pm$ 0.4 \\
Horava--Lifshitz Gravity (breaking speed of light limit at high energies) & 21 $\pm$ 4.6 & 1.3 $\pm$ 0.3 \\
Causal Dynamical Triangulations (gluing triangulated space in layers of time) & 16 $\pm$ 4.0 & 1.0 $\pm$ 0.2 \\
Asymptotic Safety (spacetime becomes a fractal on small scales) & 88 $\pm$ 9.4 & 5.3 $\pm$ 0.6 \\
Gravity is not quantum & 295 $\pm$ 17.2 & 17.7 $\pm$ 1.0 \\
Other & 192 $\pm$ 13.9 & 11.5 $\pm$ 0.8 \\
No opinion & 478 $\pm$ 21.9 & 28.7 $\pm$ 1.3 \\
\end{longtable}

\bibliographystyle{apsrev4-2}
\bibliography{aps_survey_refs}

\end{document}